11-23-2017

# The Research on the Stagnant Development of Shantou Special Economic Zone Under Reform and Opening-Up Policy

Bowen Cai

# The Research on the Stagnant Development of Shantou Special Economic Zone Under Reform and Opening-Up Policy

**Abstract**
This study briefly introduces the development of Shantou Special Economic Zone under Reform and Opening-Up Policy from 1980-2016 with a focus on policy making issues and its influences on local economy. This paper is divided into two parts, 1980-1991, 1992-2016 in accordance with the separation of the original Shantou District into three cities: Shantou, Chaozhou and Jieyang in the end of 1991. This study analyzes the policy making issues in the separation of the original Shantou District, the influences of the policy on Shantou's economy after separation, the possibility of merging the three cities into one big new economic district in the future and reasons that lead to the stagnant development of Shantou in recent 20 years. This paper uses statistical longitudinal analysis in analyzing economic problems with applications of non-parametric statistics through generalized additive model and time series forecasting methods. The paper is authored by Bowen Cai solely, who is the graduate student in the PhD program of Applied and Computational Mathematics and Statistics at the University of Notre Dame with concentration in big data analysis.

# The Research on the Stagnant Development of Shantou Special Economic Zone Under Reform and Opening-Up Policy

**Bowen Cai[1]**

1 Department of Applied and Computational Mathematics and Statistics, University of Notre Dame, Notre Dame, Indiana, U.S.A. bcai1@nd.edu

205-886-9176

**Abstract.**

This study briefly introduces the development of Shantou Special Economic Zone under Reform and Opening-Up Policy from 1980-2016 with a focus on policy making issues and its influences on local economy. This paper is divided into two parts, 1980-1991, 1992-2016 in accordance with the separation of the original Shantou District into three cities: Shantou, Chaozhou and Jieyang in the end of 1991. This study analyzes the policy making issues in the separation of the original Shantou District, the influences of the policy on Shantou's economy after separation, the possibility of merging the three cities into one big new economic district in the future and reasons that lead to the stagnant development of Shantou in recent 20 years. This paper uses statistical longitudinal analysis in analyzing economic problems with applications of non-parametric statistics through generalized additive model and time series forecasting methods. The paper is authored by Bowen Cai solely, who is the graduate student in the

PhD program of Applied and Computational Mathematics and Statistics at the University of Notre Dame with concentration in big data analysis.

**Introduction.**

After decades of centrally planned economy, Chinese central government adopted the Open Door policy in 1978. Shenzhen, Zhuhai, Shantou in Guangdong Province, and Xiamen in Fujian Province were designated as special economic zones (Zeng, 2015). These four special economic zones share a common spatial feature. That is, each special economic zone is engaged in a specific, well-targeted cooperation (Yuan, 2017). Shenzhen is adjacent to Hong Kong. Zhuhai shares the border with Macau. Xiamen is opposite to Taiwan. Shantou is faced to South East Asia, since most of Chinese immigrants in South East Asia have roots in Teochew-Shantou Area. On the one hand, the geographical and historical advantages of these four special economic zones help them attract investments from overseas Chinese and foreign countries by introducing more relaxed regulations, including export goods duty-free and reduced tax in fixed income. On the other hand, due to their adjacency to certain targeted area, they can act as the factory of their neighbors to produce goods by making use of cheaper local labor market. Some conclusions about the reforms in the special economic zones can be drawn: first, new systems involving capital and labor were established as substantive content from the perspective of the producer to break through the shackles imposed capital and labor by the old socialized planned system, and form specific subjects of rights to create an incentive for factor owners to make it possible to change the the urban economic system; second, the reform of the system of pricing plus the establishment of a market trading system enabled the trading of factors and goods (Yuan, 2017). Generally speaking, it is proved to be successful for the practice of Open Door Policy. Shenzhen is the spectacular district that benefited most from the reforms and opening-up policy. It grew up from a small fishing village to one of the biggest three cities in China with annual GDP growth rate of 40% between 1981 to 1993. GDP per capita in Shenzhen rose from RMB 606 (around US $260) in 1979 to RMB 160,000 (around US $ 25,500) in 2015 (Shen, 2008). The GDP in the city of Shenzhen was greatly expanded from RMB 196 million in 1979 to RMB 1.95 trillion in 2016 (Shen, 2008). The economy in Zhuhai and Xiamen got significantly expanded also. These two cities quickly became leading cities with strong economy in Guangdong and Fujian provinces after Open Door policy. However, Shantou, once was the second economy in Guangdong in early 1980s declined to the thirteenth economy in Guangdong in 2015. In the year of 2001, when all other cities were fast growing, Shantou experienced -1.8% (Luo, Wang & Zhong, 2007) increase in economy. The reasons for experiencing slow growth and even negative growth within the rapid growing period is multi-facets. In this paper, we are using economic and statistical approaches to explain the factors leading to this abnormal situation in Shantou. One noticeable event within the past 35 years is the separation of Shantou district into Shantou, Jieyang and Chaozhou in 1991. We will focus on the policy making issues in analyzing this separation decision and its influences on Shantou's economy after the separation. We will discuss the factors that may lead to the unsuccessful practice in Shantou special economic zone in terms of culture, investment environment, politics and others. In the end of the paper, we will discuss the possibility and necessity of combing these three cities into one municipality along with its policy making considerations. Statistical methods will be intensively applied. We will use non-parametric generalized additive model for retrospective analysis and time series model for prospective forecasting.

**Shantou's economy in the period of 1980-1991**

Shantou is a city significant in 19$^{th}$ century Chinese history as one of the treaty ports established for Western trade and contract. It locates in eastern coast of Guangdong, which is the midpoint between Hong Kong and Kaohsiung (Taiwan). Most of overseas Chinese in South East Asia have roots in Shantou-Teochew area. As an open port for more than 100 years since 1861, Shantou is a transport hub and a merchandise distribution center in Southeast China for a long time. The foundation for being special economic area is more mature than many other cities in Guangdong. Shantou Port's cargo throughput was the second in Guangdong before 1980s, and so was its GDP at that time. Due to its ideal foundation in port's economy, having millions of immigrants in overseas and special geographical location, Shantou was selected as one of the four first-tier opening cities in China in 1981. Chinese government designated Shantou to be a special economic zone with main focus on export and tourism. The administrative area of Shantou district was 10,415 square km, including areas that covered by Jieyang city and Chaozhou city nowadays (Shen 2008). Shantou Special Economic Zone is situated on a piece of sandy land in the eastern suburbs of

Shantou city, and originally covered an area of 1.6 square km in 1981 (Shen 2008). The vast inland area of Shantou was also considered to be a potential advantage over other economic zones in China, since it had inland area to rely on and to lessen city burdens through transfering manufactures. The practice of establishing special economic zone in Shantou was proved to be very successful in early 1980s with GDP annual growth rate of 12.69% (Luo, et al, 2007). Shantou special economic zone expanded its area to 52.6 square km in November 1984 (Luo et al, 2007). Started in 1985, Shantou experienced a fast growth in GDP with 28.65% annual growth rate through 1991 (Luo et al, 2007). The following graph shows the growth of Shantou's GDP in the period of 1980 -1991 (the unit of y-axis (GDP) is 100 millions RMB).

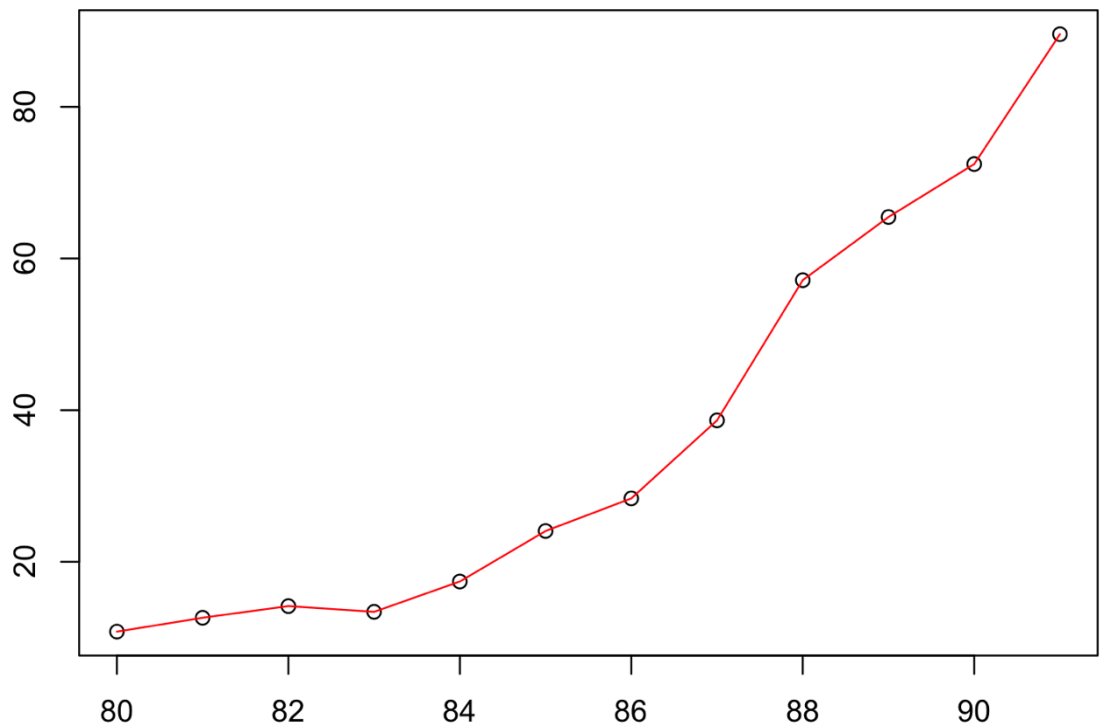

The success of Shantou's rapid growth during this 12 year-period can be ascribed to three main factors.

First, the special policy for economic zones. Shantou special economic zones enjoyed duty-free for export goods and tax reduction for import means of production. Many overseas investors chose Shantou as their priority investment place.

Second, the support of labor intensive industry from inland area. Shantou district including Jieyang and Chaozhou, had cheap labor resources. It formed three-dimension economy with cargo trades and transportation in Shantou special economic zones, labor intensive and manufacturing industry in Jieyang, and agricultural and tourism in Chaozhou. The vast land in Jieyang and Chaozhou backed up Shantou's economy and made urban area focus mainly on developing cargo trades and services.

Third, the concentration of resources. Airport, train station, major ports, municipal government, higher education institutions and enterprises were all in Shantou urban district. Jieyang and Chaozhou were Shantou's functional counties, and their main characters were to support this rapid growing economy. The highly concentrated resources stimulated and contributed to the growth of Shantou's economy.

However, in 1991, Shantou was the most populous special economic zone (SEZ) among the four cities. There were 10 million people in Shantou including Jieyang and Chaozhou. Admittedly, it created fiscal burden for the special economic zone. Some policy makers thought it would be a better idea to divide the original Shantou district into three cities, Shantou, Jieyang and Chaozhou, which could prevent resources for Shantou special economic zone from being distributed by Jieyang and Chaozhou. Another reason to separate Shantou district is because Shantou (eight counties and one city) was the largest city among four special economic zones, which did not meet the requirement that each special economic zone was only allowed to take one county as its affiliation. However, it is unfair to Shantou since the other three special economic zones were newly built cities, only Shantou is an old port with a thousand-year history. It is not wise to ask all four special zones to meet the same requirement regardless of their unique histories and conditions.

As a momentum for stimulating the development of economy, Chinese central government enlarged the special economic zone to the whole urban district in new Shantou city in compensation for the independence of Jieyang and Chaozhou. Many policy makers thought that without hindering from inland area, Shantou would get greatly developed with enjoying the special economic policies alone. However, the reality turns out to be that the sacrifice of Jieyang and Chaozhou did not bring back the blossom of Shantou's economy even though the square of special economic zones was enlarged to be four times the original square.

**Shantou's Economy in 1991-2007**

After a short period of increasing from 1992 to 1997, Shantou's economy became stagnant for the followed five years. We can see from the plot as follows.

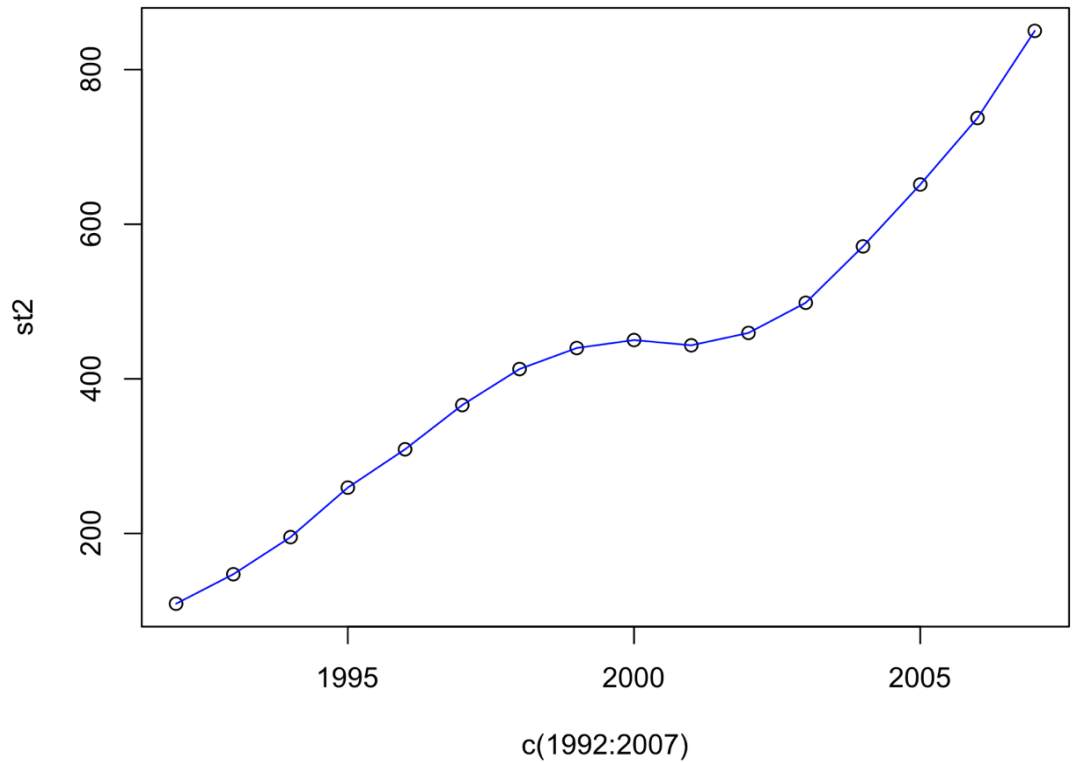

It is easy to notice that the economy expanded very little from 1997 to 2003. In 2000-2001, the economy even experienced negative growth. This five-year period however, was the rapid growth time for the rest part of China. The advantages of Shantou special economic zone rapidly disappeared after this five-year depression.

We can explain the short increasing period during 1992-1997 by using statistical approaches and economic interpretations.

Statistical Approach: Generalized Additive Model with different penalties:

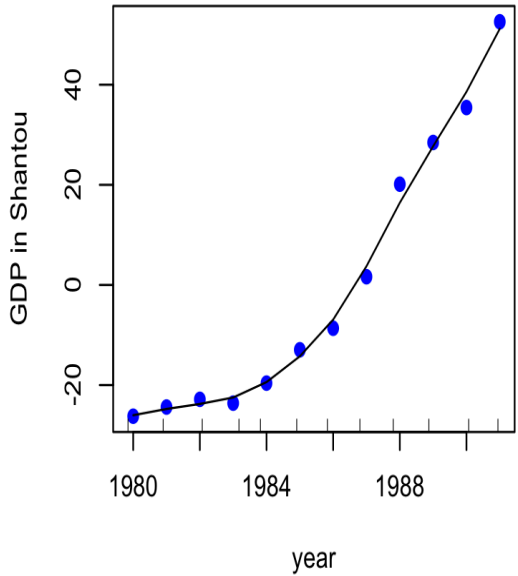
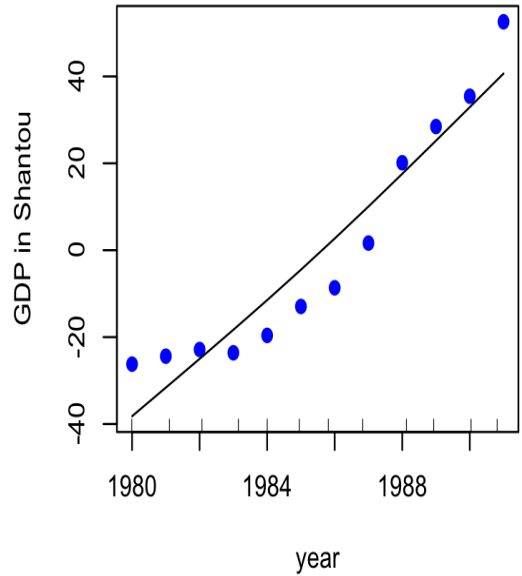

We can use smoothing splines to model GDP movements in Shantou over the period of 1980 to 1991. A key feature of smoothing splines approach is to saturate the predictor space with knots and then protect against overfitting by constraining the impact the knots can have on the fitted values (Berk, R.A. (2016), *Statistical Learning from a Regression Prospective.* Philadelphia, PA: Springer). The goal for this approach is to minimize the penalized error sum of squares of the form

$$\text{RSS}(f, \lambda) = \sum_{i=1}^{N}[y_i - f(x_i)]^2 + \lambda \int [f''(t)]^2 dt,$$

where lambda is a tuning parameter. The first term on the right-hand side captures how close the fitted values are to the actual values of y (Berk, R.A. (2016), *Statistical Learning from a Regression Prospective.* Philadelphia, PA: Springer). The second imposes a cost for the complexity of the fit, similar to the penalized regression. The larger the value of lambda, the smoother the representation of the association between X and Y. As a result, the more bias and less variance (Berk, R.A. (2016), *Statistical*

*Learning from a Regression Prospective.* Philadelphia, PA: Springer). The less the value of lambda, the more vivid the representation of the association between X and Y. It will likely to lead to less bias and more variance (Berk, R.A. (2016), *Statistical Learning from a Regression Prospective.* Philadelphia, PA: Springer). Therefore, the value of lambda can be used in place of the number of knots to tune the bias-variance tradeoff (Berk, R.A. (2016), *Statistical Learning from a Regression Prospective.* Philadelphia, PA: Springer). In our case, we use lambda to be 0.4 and 0.8 respectively. It is easy to notice that when we set tuning parameter to be 0.4, the fitting line is more vivid. When lambda is 0.8, the fitting line goes straight forward. However, no matter how large the size of penalty is, both fitting plots indicate the booming up trend will increase for the next few years. Nonparametric statistics shows that after the separation of the original Shantou district, the economy would keep increasing in a short period.

Economic factors interpretation: a straight forward interpretation of the economy blossom in the following five years after separation was because the enlarged area of special economic zone from 52 square km to 234 square km in 1992 (Luo et al, 2007). Shantou's economy benefitted a lot from the extra 182 square km special economic zone, and the original factors of production did not disappear immediately. Therefore, the economy would not recede as soon as the separation of Shantou district. However, the separation of Shantou district created potential threats to the sustainable development of the special economic zone. Fewer supply of cheap labors from inland area, lots of left over manufacturing factories in Jieyang, prosperous tourism development in Chaozhou, cutthroat competitions among three cities, distribution of overseas Teochew Chinese investment, implied that the long term sustainable and rapid development of Shantou was no longer possible. The fall down of the city was forthcoming.

The fall down of the city happened in 1998-2002. The annual GDP increase rate is 2.42% (Luo et al, 2007). In the year of 2000, the city even experienced 1.8% (Luo et al, 2007) decrease in GDP compared to 1999. The reasons that lead to the stagnant development of Shantou within this five-year period are multi-facets. Three major factors should be considered.

First, since 1980, more than 90% foreign capital in Shantou came from Asian Pacific region. More than 85% export goods from Shantou were sold to Asian countries. However, in 1998, Asian financial crisis covered most of regions in South East Asia. Countries that influenced by Asian financial crisis did not have enough capital to invest. Shantou's economy extremely relied on foreign trades and services. The depression in south east Asian countries would inevitably result in the recede of Shantou's economy. Shantou had already transferred its manufacturing factories to Jieyang, and agricultural productions to Chaozhou. The leftover in Shantou were trades and services, which were the industries that easily get impaired by the financial crisis.

Second, due to its extreme rely on trades, in late 1990s, tax evasion and smuggling were universal in Shantou, which ruined its recognition and city image to investors. Fewer investments were introduced to Shantou compared to previous years. As a *New York times* reporter writes, "In places such as Shantou, where the smuggling of diesel oil, cars, cell phones, and much more has long underpinned the economy" (Rosenthal, "Smuggling War Rippling through China", *New York Times*, March 6, 2000). The use of faked export of receipts in Shantou allowed local companies to claim central government tax rebates of more than $500 million without really export goods (Cartledge, "China's Leaders ignore Graft at Their Peril", *New York Times*, July 6, 2003) . The city corruption in late 1990 and early 2000 was one of the main reasons that led to the decrease of economy.

Third, the development of Jieyang and Chaozhou distributed the limited resources. Separation of Shantou district did not meet policy makers' expectation of concentrating resources, rather it split the already limited resources to three prefecture cities. Although most of overseas immigrants have roots in Teochew, they preferred making investments in their hometown, rather than Shantou itself. Formerly, when Jieyang and Chaozhou belonged to Shantou district, there was no difference for them to choose which city to invest. However, after separation, they hoped to help root place local economy develop rather than merely invest in Shantou to get the benefit of tax reduction. The competition from Jieyang and Chaozhou suppressed the development of Shantou as well. There is no wonder that after a short period of spring, Shantou experienced chilly winter immediately. Until 2003, Shantou's economy started to recover from depression.

Let's fit a non-parametric generalized additive regr0ession line for Shantou's economy from 1980 to 2007.

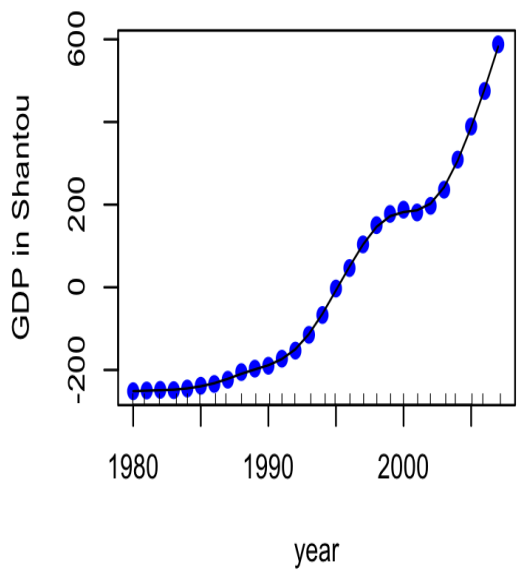 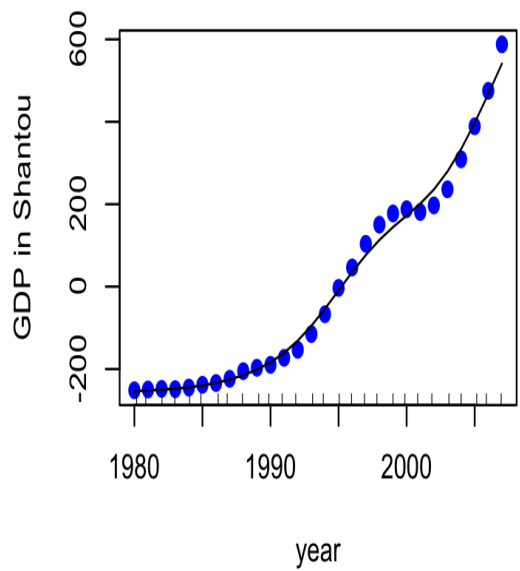

For penalty of 0.4, we see that the regression line models the trend vividly. For penalty of 0.6, we see the general trend went up for the 28-year period of 1980-2007. Admittedly, reform and opening-up policy contributed to the development of Shantou's economy significantly, even though the growing up speed was not ideal and Shantou lost its influence in Guangdong by falling from the second place in 1980 to the eleventh place in the end of 2007 in terms of GDP. In 1989, Shantou district was the third strongest economy in Guangdong with GDP of 101,500 millions RMB, while Shenzhen was one place behind Shantou with GDP of 97,800 millions RMB (Guangdong Government Bureau of Statistics, 1990. *Statistics Yearly Report*.). However, in 2007, Shenzhen's GDP is 8 times of Shantou. Part of the reason was due to the separation of Shantou district. Should we encourage communications among Shantou, Jieyang and Chaozhou? Is there any possibility for these three cities merging into greater Shantou again?

**Discussion on the pros and cons of merging three cities into one greater district**

Pros for merging the three cities together:

These three cities share same history and culture. The three cities were one integral in the last one thousand years. Divided into three parts leads to geographical segregation among three prefectures. As a result, some issues pop out.

First, waste of resources. Redundant projects are universal. Three prefectural governments exist at the same time on the area that one prefectural government would be enough for the administration. The existence of three municipal government increases fiscal burden for local economy.

Second, exacerbate competition for limited resources. The three cities were too close to each other. The division of this integral lead to cutthroat competition for limited resources. For example, airport, train station should all be built on the midpoint of three cities. Shantou is the smallest among these three cities, and is far away from the midpoint. Therefore, airport was built in Jieyang, and train station was built in Chaozhou. Shantou became the only special economic zone without airport and train station. Furthermore, the distribution of limited resources makes it impossible for any one of them to take big construction project, which is hard to form leading industry.

Third, the functions of Shantou special economic zone were largely compromised. Shantou once was the center city of east Guangdong. However, nowadays, Jieyang and Chaozhou competed for being the leader of east Guangdong as well. Multi-centers are equivalent to no centers. The economic status of Shantou is greatly reduced. The special economic zone cannot perform well as expected to improve surrounding economy because its functions were largely compromised by neighbors. Shantou has no advantages over Jieyang and Chaozhou except for policies. Nevertheless, the whole China is open to the world, previous open door policies in special economic zone is becoming less important.

Therefore, merging three cities together can stand out Shantou's economy in leading the whole development of Teochew-Shantou area rather than competing with each other for limited resources.

Cons for merging the three cities together

Jieyang and Chaozhou have been independent from Shantou district for twenty years. They have formed respective regulations and administrative policies. They have their own transportations, bureaus and departments. If three cities merged together, then a series of issues including human resource, energy, administration need to be carefully considered. Although competitions among three cities distribute limited resources, we are glad to see that Jieyang's economy surpassed Shantou in 2015. In 2015, Jieyang's GDP reached 189,001 million RMB, while Shantou's GDP was 186,803 millions RMB (Shen 2008). Regional competition also improves regional economy. It would be better to encourage these three cities to communicate more with each other, rather than forcing them into one administrative region. Resources may be concentrated by merging them into one city, but respective advantages may lose as well. Increasing communication and cooperation among three cities can be supplement to forcing them being one administrative area. In recent five years, joint conferences held by mayors of three cities every year. A range of priorities and specific projects have been identified to support cooperation, covering the realms of economic and industrial development, transport, infrastructure, telecommunication, television networks, healthcare, financial settlement, human resource management, policing and mobile population regulation, water management and so on so forth (Li, Wu & Hay, 2015). Instead of administrative coalescence into one municipality, Shantou-Jieyang-Chaozhou could attempt to promote integration and cooperation by means of a flexible governance framework at the city-regional scale (Li et al, 2015). They could start with coping with environmental pollution and intercity transportation together, and then discussed the possibility and necessity of coalescing into one city. Furthermore, whilst Shantou-Jeiyang-Chaozhou integration has proved to be a simple inclusion in provincial government documents, it has not been matched with substantial, tangible support from that provincial government (Li et al, 2015). It may be more emergent for them to tackle with the issues of competing with each other, ambitious to develop into a region's leading city and hinterland the development of the others than merely waiting for the decision of central and provincial government merging them into one municipality.

**Time Series forecasting for Shantou's GDP in the next few years**

Shantou caught up after 2003, with annual GDP increase rate of around 10%. In 2016, Shantou achieved the breakthrough for the first time to break 200,000 millions RMB in its history. In the first quarter of 2017, Shantou's GDP growth rate is the highest among all cities in Guangdong. We can see the booming up trend in the following graph

from 2008-2016.

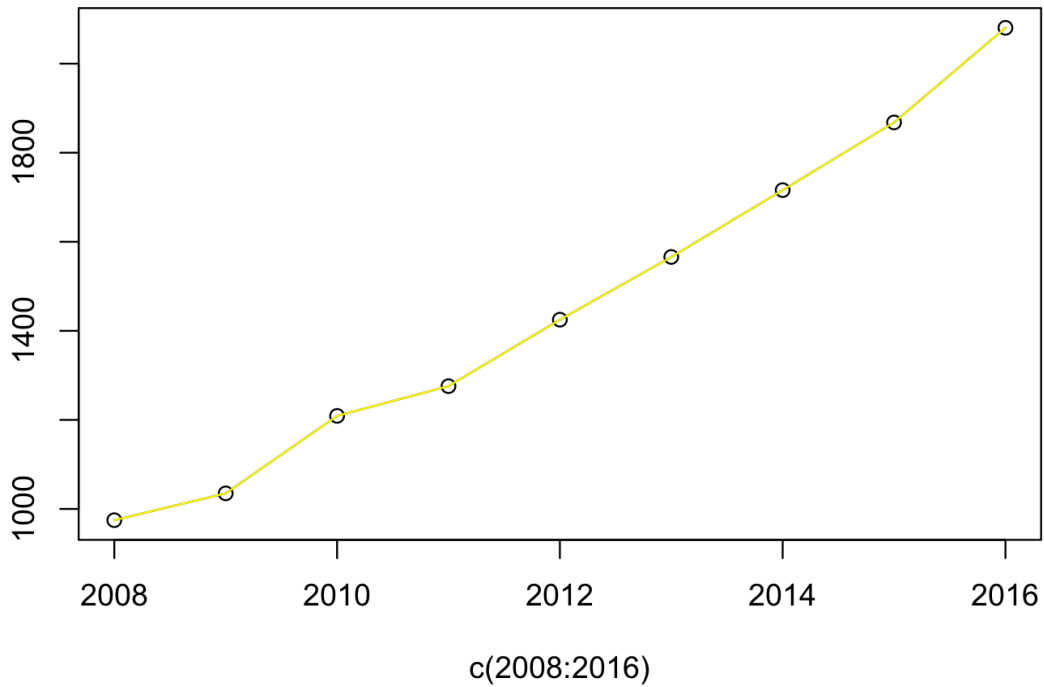

Here, we want to introduce the statistical methods to predict the total volume of Shantou's GDP in the next few years. One common statistical method is time series. A time series is a discrete time, continuous state process. A sequence of random variables indexed by time is called a stochastic process (stochastic means random) or time series (Imdadullah. "Time Series Analysis". *Basic Statistics and Data Analysis*. itfeature.com. Retrieved 2 January 2014). A data set is one possible outcome of the stochastic process. If history had been different, we would observe a different outcome, thus we can think of time series as the outcome of a random variable (Imdadullah. "Time Series Analysis". *Basic Statistics and Data Analysis*. itfeature.com. Retrieved 2 January 2014). Rather than dealing with individuals as units, the unit in time series forecasting methods is time. The value of Y at time t is $Y_t$. The unit of time in our analysis is year. The value of $Y_t$ in the previous period is the first lag value: $Y_{t-1}$. The jth lag is denoted as $Y_{t-j}$. $Y_{t+1}$ is the value of $Y_t$ in the next period (Imdadullah. "Time Series Analysis". *Basic Statistics and Data Analysis*. itfeature.com. Retrieved 2 January 2014). In our case, we fit a regression based time series forecasting model to the time power of 4.

```
Call:
lm(formula = stn ~ 0 + time + I(time^2) + I(time^3) + I(time^4))

Coefficients:
     time  I(time^2)  I(time^3)  I(time^4)
-9.620999   2.631056  -0.105464   0.002233
```

The summary table is displayed as follows.

```
Call:
lm(formula = stn ~ 0 + time + I(time^2) + I(time^3) + I(time^4))

Residuals:
    Min      1Q  Median      3Q     Max
-68.941 -24.823  -0.339  22.010  78.081

Coefficients:
            Estimate Std. Error t value Pr(>|t|)
time       -9.6209990  5.6542323  -1.702 0.098244 .
I(time^2)   2.6310563  0.8185008   3.214 0.002919 **
I(time^3)  -0.1054636  0.0369321  -2.856 0.007373 **
I(time^4)   0.0022326  0.0005216   4.280 0.000151 ***
---
Signif. codes:  0 '***' 0.001 '**' 0.01 '*' 0.05 '.' 0.1 ' ' 1

Residual standard error: 37.41 on 33 degrees of freedom
Multiple R-squared:  0.9981,	Adjusted R-squared:  0.9978
F-statistic:  4277 on 4 and 33 DF,  p-value: < 2.2e-16
```

We also tried to include $I(time^5)$ into the model, but it is not significant. Therefore, we only include $I(time)$, $I(time^2)$, $I(time^3)$, and $I(time^4)$ into our model to fit the regression line.

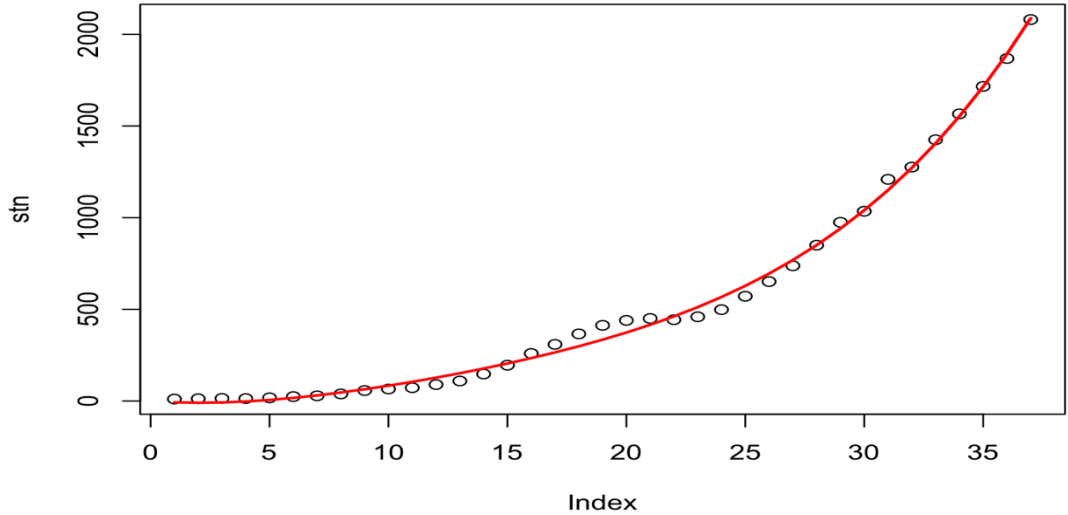

We see that even though it is to the power of four, the time series regression line is smoother than generalized additive model. It indicates that the economy will keep growing after 2016. However, the regression based time series model does not fit the trend well. We can tell this by looking at the autocorrelation plot and partial autocorrelation plot.

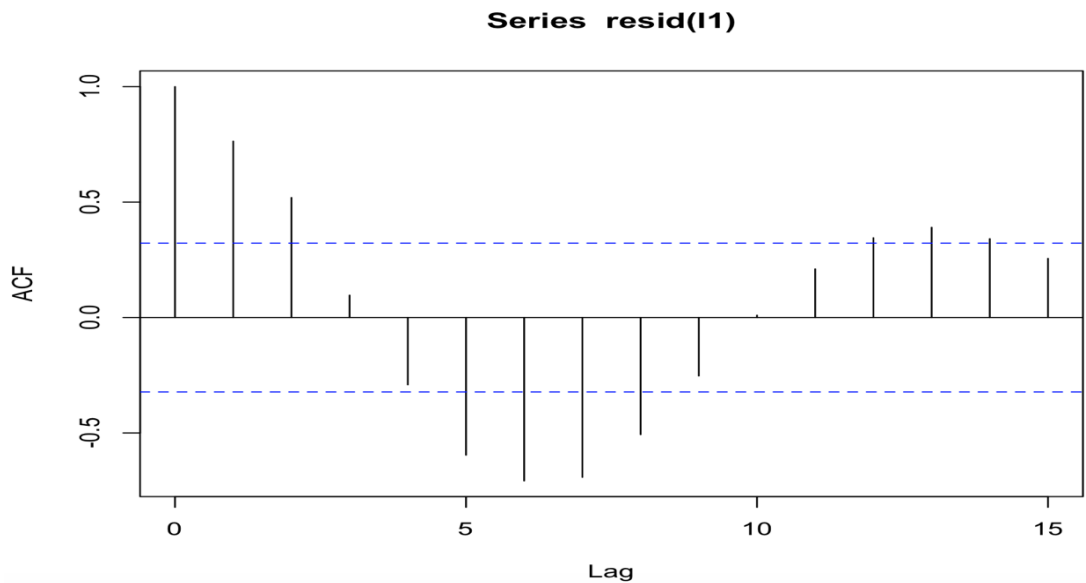

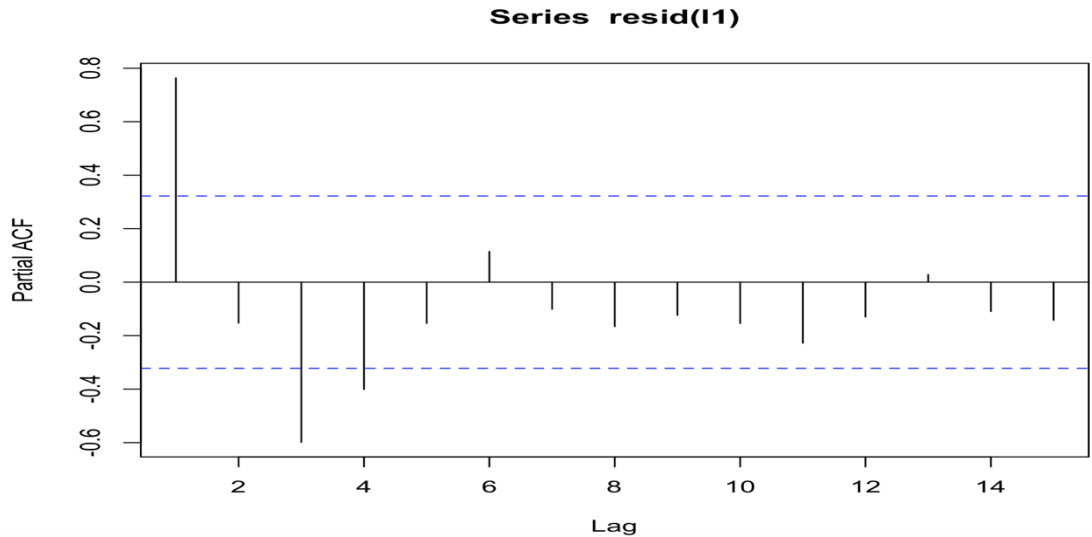

There are some spikes that it cannot capture, which means the model is not reduced to white noise. However, it is good for us to look at the predicted values by using this simple linear regression model.

```
> forecast(fitted(l1))
   Point Forecast    Lo 80    Hi 80    Lo 95    Hi 95
38       2283.337 2273.248 2293.425 2267.908 2298.766
39       2478.577 2456.022 2501.131 2444.083 2513.071
40       2673.817 2636.078 2711.556 2616.101 2731.534
41       2869.057 2813.815 2924.300 2784.571 2953.544
42       3064.298 2989.500 3139.095 2949.905 3178.691
43       3259.538 3163.328 3355.748 3112.397 3406.679
44       3454.778 3335.445 3574.112 3272.274 3637.283
45       3650.019 3505.970 3794.068 3429.715 3870.322
46       3845.259 3674.998 4015.520 3584.867 4105.651
47       4040.499 3842.610 4238.388 3737.854 4343.144
```

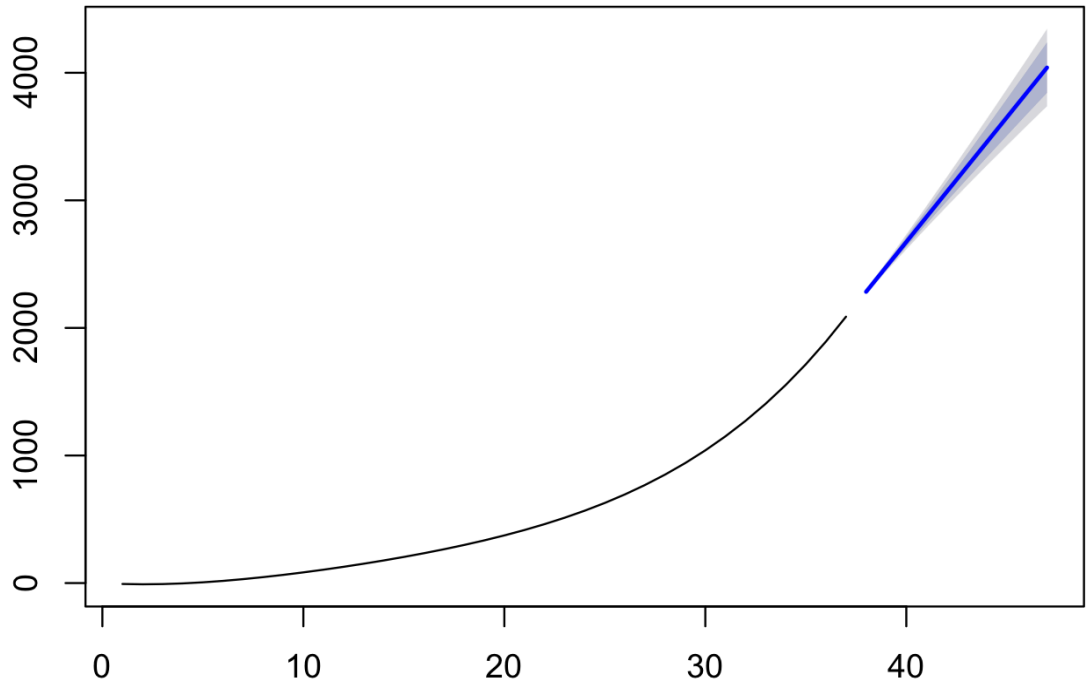

```
     Point Forecast      Lo 80     Hi 80     Lo 95     Hi 95
38        2283.337   2273.248  2293.425  2267.908  2298.766
39        2478.577   2456.022  2501.131  2444.083  2513.071
40        2673.817   2636.078  2711.556  2616.101  2731.534
41        2869.057   2813.815  2924.300  2784.571  2953.544
42        3064.298   2989.500  3139.095  2949.905  3178.691
43        3259.538   3163.328  3355.748  3112.397  3406.679
44        3454.778   3335.445  3574.112  3272.274  3637.283
45        3650.019   3505.970  3794.068  3429.715  3870.322
46        3845.259   3674.998  4015.520  3584.867  4105.651
47        4040.499   3842.610  4238.388  3737.854  4343.144
```

We see that it predicts in the year 2017, GDP in Shantou will be 2283.337 hundred of millions RMB, and in 2018, it will be 2478.477 RMB. If we calculate the growth rate, we will get GDP growth rate for 2016-2017 is 9.74%, while GDP growth rate for 2017-2018 is 8.55% as predicted by this model. It shows a downturn in the future. It is reasonable, because it is always hard to maintain high increasing speed without outside stimuli. We can check the accuracy of this forecasting by using the first half year data in 2017 to estimate Shantou's economy in the end of the year. Shantou's GDP in the first half year is 1064.87 hundred of millions RMB. We can estimate the total volume of Shantou's GDP in the end of the year if the growth rate is constant, and we get the whole year GDP will be around 2269.869 hundred of millions RMB. It seems like the regression model over-estimates Shantou's economy performance in 2017. Let's fit another model.

We can use ARIMA (0,2,2) model to capture features of this trend. We express this model as $\hat{Y}_t = 2Y_{t-1} - Y_{t-2} - \theta_1 e_{t-1} + \theta_2 e_{t-2}$, where $\theta_1$ and $\theta_2$ are MA(1) and MA(2) coefficients. This is a general exponential smoothing model. It uses exponentially weighted moving averages to estimate both a local level and a local trend in the series. We can check autocorrelation and partial autocorrelation graphs for this model (Nau., R. (2017). Statistical Forecasting: notes on regression and time series analysis. Retrieved from https://people.duke.edu/~rnau/411home.htm).

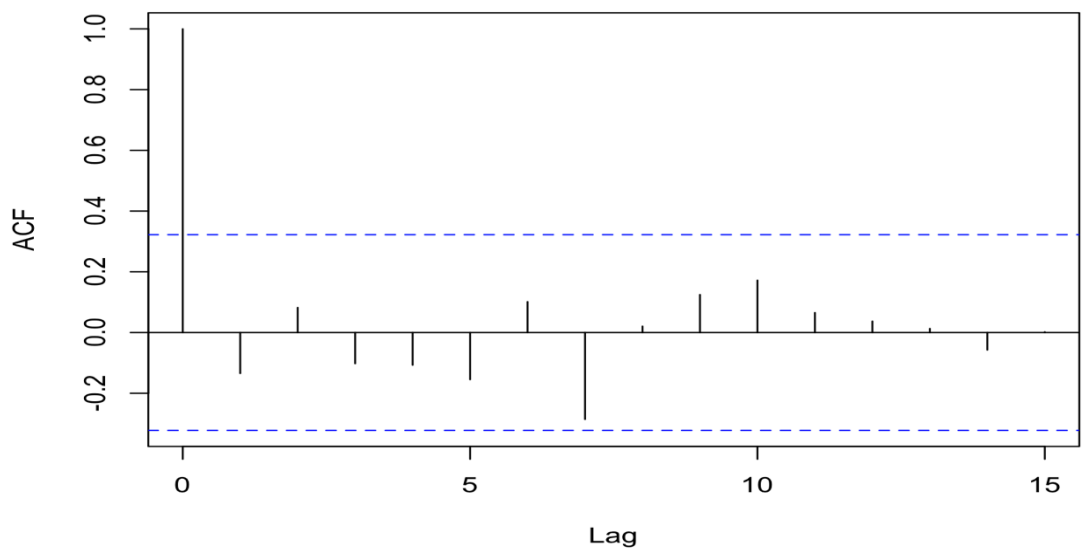

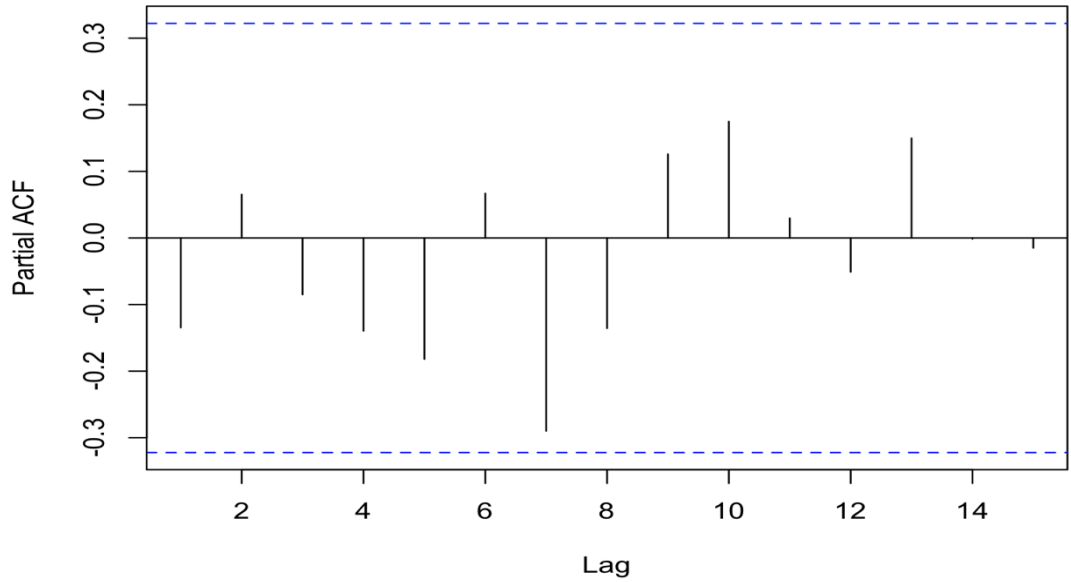

Apparently, this model is reduced to white noise, which is very ideal. If we use this model to fit the past 37 years GDP, then we can get the following graph.

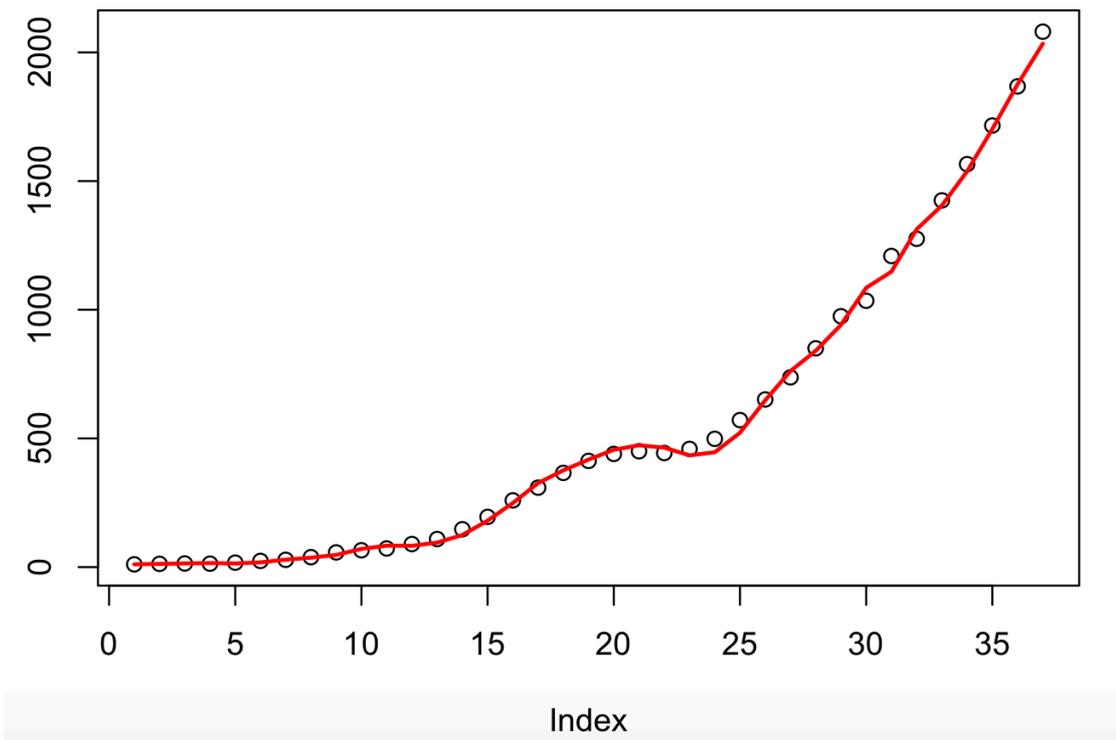

Compared to the previous regression based time series model, it fits the points more accurate. Then, let's see the prediction for the next few years by applying ARIMA (0,2,2) model.

```
> fcast
   Point Forecast    Lo 80    Hi 80    Lo 95    Hi 95
38        2259.066 2226.078 2292.054 2208.616 2309.516
39        2468.133 2411.707 2524.558 2381.837 2554.428
40        2677.200 2579.187 2775.212 2527.303 2827.097
41        2886.267 2735.544 3036.990 2655.756 3116.778
42        3095.333 2883.488 3307.179 2771.344 3419.323
43        3304.400 3024.309 3584.492 2876.038 3732.763
44        3513.467 3158.789 3868.146 2971.033 4055.901
45        3722.534 3287.474 4157.594 3057.167 4387.901
46        3931.601 3410.780 4452.421 3135.075 4728.127
47        4140.668 3529.040 4752.296 3205.264 5076.072
```

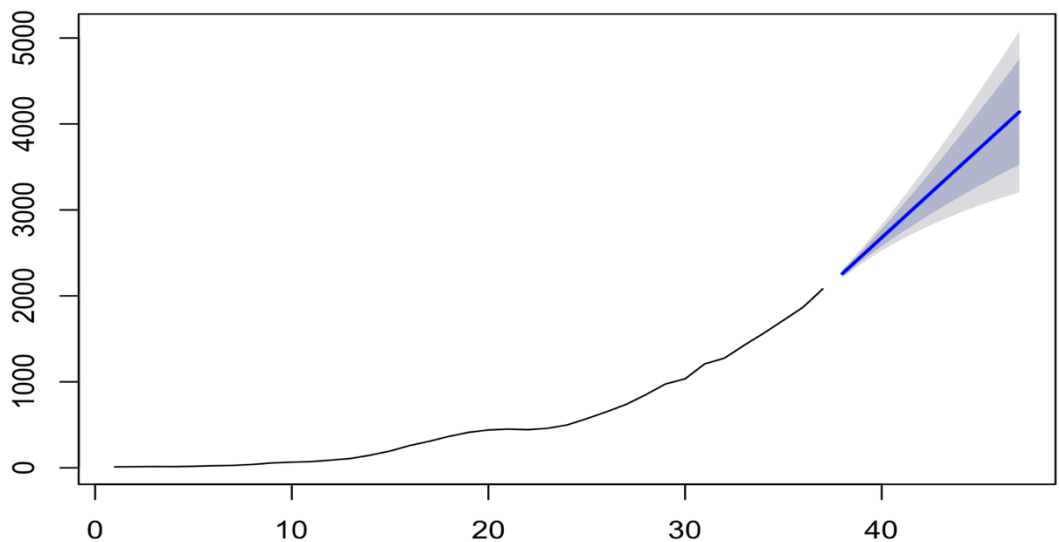

```
> fcast
   Point Forecast    Lo 80    Hi 80    Lo 95    Hi 95
38        2259.066 2226.078 2292.054 2208.616 2309.516
39        2468.133 2411.707 2524.558 2381.837 2554.428
40        2677.200 2579.187 2775.212 2527.303 2827.097
41        2886.267 2735.544 3036.990 2655.756 3116.778
42        3095.333 2883.488 3307.179 2771.344 3419.323
43        3304.400 3024.309 3584.492 2876.038 3732.763
44        3513.467 3158.789 3868.146 2971.033 4055.901
45        3722.534 3287.474 4157.594 3057.167 4387.901
46        3931.601 3410.780 4452.421 3135.075 4728.127
47        4140.668 3529.040 4752.296 3205.264 5076.072
```

It is not hard to find that this model is a little more conservative than regression based time series model. It forecasts the GDP for 2017 is 2259 hundred of millions RMB, while the regression model forecasts the volume to be 2283 hundred of millions RMB. The estimation for 2017 by using the half year statistics is 2269 hundred of millions RMB, which is in between. Although we will not be able to check the accuracy for these two models until next year, we think that the real situation will not vary much from these two prediction values. Apparently, 2269 is within the 80 % and 95% confidence interval for both of the regression based time series and ARIMA (0,2,2) models.

**Conclusion.**

Shantou was one of the earliest opening-up cities in China. For more than 120 years, over 1860s-1980s, Shantou's economy was the second strongest in Guangdong province. However, since 1990, Shantou's economic status was gradually superceded by Foshan, Shenzhen, Dongguan and Zhuhai. In 2010, Shantou's economy fell down to the twelfth place in Guangdong. Various reasons can explain this long term depression of Shantou special economic zone. Three most crucial factors are as follows.

First, the division of original Shantou district. Shantou-Jiyang-Chaozhou was one integral in the history for almost one thousand years. They share the same culture, same history, and same economy. They rely on each other. The rapid development of Shantou district in 1980s was credited to the holistic contribution for all three regions. However, the separation of Shantou-Teochew area in 1991 not only split the limited resources, but increased cutthroat competition as well. Shantou became a city heavily relied on services and foreign investments. Without support from Jieyang and Chaozhou, it is hard for Shantou to develop standing out industry within a short time.

Second, the corruption of Shantou special economic zone. The use of fake export receipt, smuggling, tax dodging and tax evasion ruined the city image in both domestic and foreign investors' eyes. Some legal and well-disciplined enterprises in Shantou also got influenced by corruption of the city. Lack of credence among investors creates hardship for further development of the city.

Third, the unique culture of Teochew area. Shantou, Jieyang and Chaozhou were all belong to Teochew area. Some traits of this area are conservative, feudal and business money goer. They are conservative because they resist to outside affairs and new residents. They speak their own language, which is much different from Mandarin and Cantonese. They like to maintain their own lifestyles without being changed by the outside world. The other three special economic zones were newly built cities. They do not have ingrained culture and history. Therefore, they focus more on developing economy and are open to reforms. Teochew people are feudal because they think boys have higher status than girls. Usually, they will not ask daughters to go to colleges, and will ask sons to take over their business in a young age. More and more less educated people can also be the reason that Shantou's economy is lack of momentum in recent years. Teochew people are good at business. However, they treat everything as business. They value too much on personal relationships rather than regulations. They are extreme money-goers. They think money can solve everything, even for things that are against laws. The unique culture in Shantou predestinates the depression of Shantou's economy even though they have the best policy.

Life is a long march. However, the success of a person is usually determined by a very few steps at certain time points. Human beings is like this, and so is the city. Many reasons can explain the depression of economy in Shantou, while the simplest reason is that this city went the wrong way for a few but crucial steps.

**References.**